\documentclass{article}
\usepackage[utf8]{inputenc}
\usepackage[lmargin=0.65in,rmargin=0.65in,tmargin=0.9in,bmargin=0.75in]{geometry}
\usepackage{hyperref,lmodern,textcomp,graphicx}
\usepackage{enumitem}
\setlist{nolistsep}
\setlist[itemize]{leftmargin=*}
\usepackage[table]{xcolor} \usepackage{caption,fancyhdr}
\usepackage{color,amsmath,amssymb,mathpazo,mathtools}
\usepackage{vwcol}
\usepackage{lipsum}
\usepackage{tgpagella}
\usepackage{indentfirst}
\usepackage{bchart}
\usepackage{lettrine,multicol}
\usepackage[sorting=none,style=chem-angew,autocite= superscript]{biblatex}
\usepackage{tcolorbox}%
\definecolor{mycolor}{rgb}{0.122, 0.435, 0.698}% Rule colour
\makeatletter
\newcommand{\mybox}[1]{%
  \setbox0=\hbox{#1}%
  \setlength{\@tempdima}{\dimexpr\wd0+13pt}%
  \begin{tcolorbox}[boxrule=0pt,arc=0pt,
      left=6pt,right=6pt,top=6pt,bottom=6pt,boxsep=1pt,width=\textwidth]
    #1
  \end{tcolorbox}
}

\fancypagestyle{firstpage}
{\fancyhead[L]{\textsc {\color {gray} {\large }}}    
\fancyhead[R]{\large \textsc{\color {gray} {Research Article}}}}

\newcommand{\HA}[1]{{\color{black} #1}}
 \newcommand{\HAB}[1]{{\color{black} #1}}
 
\graphicspath{{Figures/}}

\begin{document}

\thispagestyle{firstpage}
{\noindent \LARGE \textit{Zero-frequency corner modes in mechanical graphene}} \\ [1em] 
\noindent {\large \textit{Hasan B. Al Ba'ba'a}} \\

\noindent \begin{tabular}{c >{\arraybackslash}m{6in}}
    \includegraphics[]{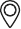} &
    \noindent {Department of Mechanical Engineering, Union College, Schenectady, NY 12308, USA} \\[0.25em]
    
    \includegraphics[]{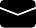}& \noindent {\href{mailto:albabaah@union.edu}{albabaah@union.edu}}\\
    
\end{tabular}

%%%%%%%%%%%%%%%%%%%%%%%%%%%%%%%%%%%%%%%%%%%%%%%%%%%%%%%%%%%%%%%%%%%%%%%%%%%%%%%%%%%%%%%%%%%%%%%%%%%%%%%%%%%%%%%%%%%%%%%%%%%%%%%%%%%%%%%%%%

\mybox{\textit{In an unconstrained elastic body, emergence of zero natural frequencies is an expectable outcome on account of the body's ability to purely translate or rotate with no structural deformation.~Recent advances in literature have pushed such conventional definition and demonstrated properties transcending typical zero-frequency modes, such as localization of deformation at a structural edge or corner.~In this paper, a spring-mass honeycomb lattice with an elastic foundation, referred to here as mechanical graphene, is designed to exhibit zero-frequency corner modes.~A central element in the proposed design is the elastic foundation, and the zero-frequency corner modes are enabled by intricate modulation of the elastic-foundation's stiffness.~These modes are proven to have their origins from the dynamics of a diatomic chain, made from a single strip of the mechanical graphene with free boundaries.~Different shapes of finite mechanical graphene with free boundaries are considered and conditions leading to the manifestation of corner modes are correlated with the angle of corners and stiffness of springs supporting them.~Finally, the effect of defects on zero-frequency corner modes is briefly discussed, demonstrating robustness against structural defects that are distant from corners.}}

\noindent\rule{7.2in}{0.5pt}

\begin{center}
\begin{tabular}{p{1.65 in} | p{5.2 in}} 

\textbf{Keywords} 
\begin{itemize}
    \item[--] Corner modes
    \item[--] Edge modes
    \item[--] Diatomic lattices
    \item[--] Mechanical graphene
    \item[--] Elastic foundation
    \item[--] Zero frequency
\end{itemize} \vfill & \textbf{Graphical Abstract}

\vspace{0.25cm} 
\includegraphics[]{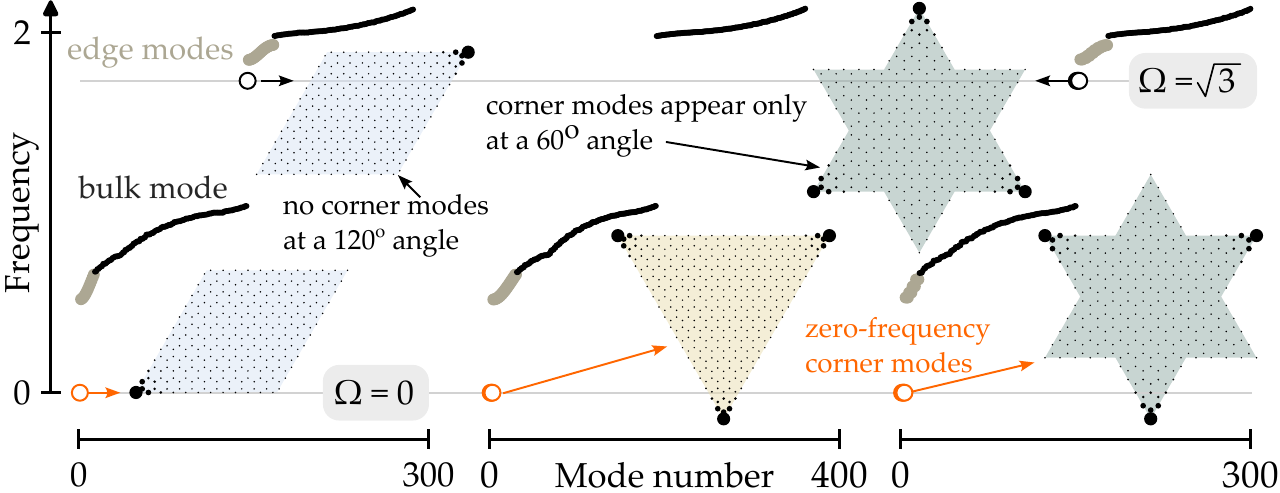} \\

\end{tabular}
\end{center}

\noindent\rule{7.2in}{0.5pt}

\section{Introduction}

Periodic lattices have been pivotal in realizing unconventional and counter-intuitive vibration phenomena in recent years \cite{Hussein2014,wang2020tunable,nassar2020nonreciprocity}. A core aspect of understanding a vibrational system is examining its natural frequencies and corresponding deformation patterns (i.e., mode shapes).~In simple rods with homogeneous elastic and inertial properties carrying longitudinal waves, the natural frequencies are often spaced evenly with no sudden increase/decrease between their individual values.~Peculiarly, the introduction of a periodic element into the structural design forces the natural-frequency spectra to separate into groups with relatively large ranges.~Having separation between these groups is now understood to be the cause of initiating bandgaps --- frequency regions of effective vibration attenuation~\cite{al2017formation,al2017pole,AlBabaa2019DispersionCrystals}. Under certain circumstances, including \HA{truncation (or termination) effects \cite{al2017pole,AlBabaa2019DispersionCrystals,davis2011analysis,al2023theory,rosa2023material}}, boundary-condition alteration \cite{bastawrous2022closed}, or defects \cite{khelif2004guided,jo2022revealing}, natural frequencies may emerge within bandgaps, enabling curious behavior combining the tendency of the vibrational system to resonate and attenuate waves concurrently. This complex interplay of two opposite phenomena results in mode shapes that of high resonant amplitudes at an edge (or a corner) of the vibrational structure, and diminishes to zero away from the edge (corner), thanks to the attenuation effects introduced by bandgaps \cite{al2017pole,AlBabaa2019DispersionCrystals,davis2011analysis, bastawrous2022closed, danawe2021existence}. These special natural frequencies have been recently explored by various researchers due to their intriguing behavior and potential in enabling a variety of applications (e.g., wave guiding \cite{addouche2014subwavelength}, \HA{flow control \cite{hussein2015flow}}, and energy harvesting \cite{lv2019highly,jo2020elastic,lee2022piezoelectric}), while having an additional layer of robustness if topologically protected \cite{susstrunk2015observation,Nash2015TopologicalMetamaterials,Serra-Garcia2018ObservationInsulator,chen2018topological,Chen2019TopologicalLattice,Fan2019ElasticStates,peterson2020fractional,wang2021elastic,an2022second}.

Being mostly finite in magnitude, natural frequencies may also assume the value of zero, particularly occurring in elastic bodies that are unconstrained.~Zero-frequency modes are often associated with ``rigid-body" motion, such that the structure would purely translate or rotate \textit{without} structural deformation.~However, certain types of lattices may exhibit zero-frequency (or near zero-frequency) modes showing wave phenomenon beyond simple rigid body modes (e.g., localization at edges), owing to a non-zero wavenumber (i.e., spatial wave frequency) as demonstrated in granular media \cite{zheng2016zero,zheng2017zero}.~Further, significant body of research in the context of Maxwell lattices \cite{maxwell1864calculation} have demonstrated the existence of topological zero-frequency floppy modes with localization at lattice's edges  \cite{kane2014topological,lubensky2015phonons,zhou2018topological,zhou2019topological}. Topologically-protected zero-frequency corner modes have been also achieved via a checkerboard pattern of interconnected rigid quadrilaterals, assembled via unconstrained hinges \cite{saremi2018controlling}. Nonetheless, zero-frequency corner modes in mechanical systems remain largely unexplored in literature, and this paper aims to address such shortcomings.

\HA{In this effort, an elastically-supported spring-mass honeycomb lattice is designed to exhibit zero-frequency corner modes (Figure~\ref{fig:DL_Schematics}(a,b)).~Termed \textit{mechanical graphene}, analogous to similar mechanical systems studied in literature \cite{socolar2017mechanical,kariyado2015manipulation}, a key element in its design is the elastic foundation, which, if carefully modulated, can yield zero-frequency corner modes.~In particular, one of the elastic-foundation} supports in the two sites of mechanical-graphene's unit cell must possess a negative spring constant, interestingly enabled without losing dynamical stability. Emergent zero-frequency corner modes stem from the dynamics of a single strip with open boundary conditions, constituting a diatomic lattice. Having diatomic lattices of either odd or even number of degrees of freedom ultimately affects the natural-frequency spectra, including modes originating within bandgaps.~Existence of zero-frequency corner modes is quantified analytically and correlated to the shape of corners and their elastic-support stiffness. Different shapes of finite mechanical graphene, shown in Figure~\ref{fig:DL_Schematics}(c), have been tested and their robustness against defects is briefly discussed.

\section{Diatomic lattice}

\subsection{Wave dispersion \HA{analysis}}
\label{sec:DL_UC}
\begin{figure*}[]
     \centering
\includegraphics[width=\textwidth]{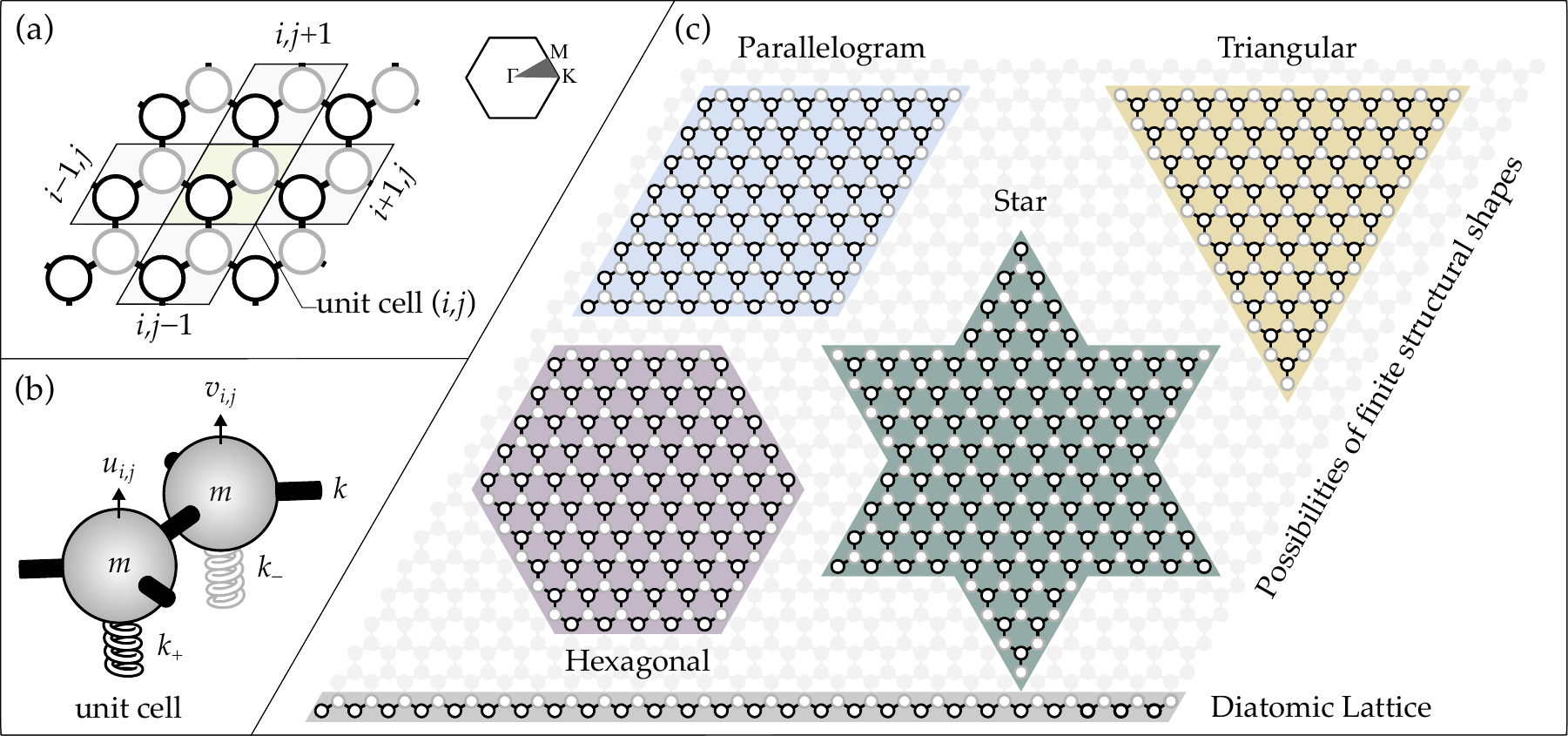}
\caption{(a) Schematics of elastically-supported mechanical graphene, defining the unit cell and illustrating its periodicity in two directions via the indices $i$~and~$j$.~The irreducible Brillouin zone is also shown for reference. (b) Close-up of the $(i,j)^{\text{th}}$ unit cell, showing its degrees of freedom $u_{i,j}$ and $v_{i,j}$ (quantifying out-of-plane displacement) as well as its inertial and elastic properties.~Taut strings with transverse stiffness $k$ couple the honeycomb array of identical masses~$m$, while $k_+$ ($k_-$) elastically-supports the masses of displacement $u_{i,j}$ ($v_{i,j}$).~(c) Depiction of different possibilities of finite mechanical graphene overall geometries, including parallelogram, triangular, star, and hexagonal shapes. Note that a single-row strip of mechanical graphene constitutes a diatomic lattice.~All cases have free boundary conditions at all of their edges.}
     \label{fig:DL_Schematics}
\end{figure*}

Zero-frequency corner modes in mechanical graphene have their roots in the dynamics of a single strip with free (unconstrained) boundaries, i.e, a diatomic lattice (See Figure~\ref{fig:DL_Schematics}\HA{(b,c)}). As such, the dynamics of a diatomic unit cell \HA{is first considered}, which has equations of motion given by~\cite{al2017pole,al2020uncertainty}:
\begin{subequations}
\begin{equation}
     m \Ddot{u}_{i,j}  + (2k+k_+) u_{i,j}  - k (v_{i,j}  + v_{i-1,j}) = 0,
\end{equation}
   \begin{equation}
     m \Ddot{v}_{i,j}  + (2k+k_-) v_{i,j}  - k (u_{i,j} + u_{i+1,j}) = 0.
\end{equation}
\label{eq:EOM_diatomic_UC}
\end{subequations}
\HA{where the springs of transverse stiffness constant $k$ couple the identical array of masses~$m$, while $k_+$ ($k_-$) elastically-supports the mechanical graphene sites with out-of-plane displacement $u_{i,j}$ ($v_{i,j}$).} To simplify analysis and reduce the number of variables, the elastic supports are parameterized as $k_{\pm} = 2k(1\pm \vartheta)$, where $\vartheta$ is a stiffness contrast variable, and the excitation frequency $\omega$ is normalized using frequency $\omega_0 = \sqrt{{2k}/{m}}$, yielding a non-dimensional frequency $\Omega~=~\omega/\omega_0$. Note that $k_+$ and $k_-$ interchange if the sign of $\vartheta$ flips, where the latter can assume any value within the range $\vartheta \in [-\sqrt{3},+\sqrt{3}]$, as will be explained shortly. Applying the Bloch-wave solution, Equation~(\ref{eq:EOM_diatomic_UC}) transforms into the following eigenvalue problem
\begin{equation}
    \mathbf{H}\mathbf{z}_{i,j}= \Omega^2 \mathbf{z}_{i,j},
    \label{eq:EVP_DL}
\end{equation}
with Hamiltonian $\mathbf{H}$ (generally a function of $x$- and $y$-direction non-dimensional wavenumbers, $q_x$ and $q_y$) and displacement vector $\mathbf{z}_{i,j}$, respectively, being 
\begin{subequations}
\begin{equation}
    \mathbf{H} = 
    \begin{bmatrix}
      2+\vartheta & -\frac{1}{2} \left(1+\text{e}^{-\mathbf{i}q_x} \right) \\
      -\frac{1}{2} \left(1+\text{e}^{\mathbf{i}q_x} \right) & 2-\vartheta
    \end{bmatrix},
\end{equation}
\begin{equation}
    \mathbf{z}_{i,j}^\text{T}= \{ u_{i,j} \ v_{i,j}\}.
\end{equation}
\label{eq:H_DL}
\end{subequations}
Here, $\mathbf{i}=\sqrt{-1}$ is the imaginary unit. Solving the eigenvalue problem in \HA{Equation~}(\ref{eq:EVP_DL}), it immediately follows that the eigenvalues of $\mathbf{H}$ are the solutions of the characteristic equation (i.e., the dispersion relation)
\begin{equation}
    \Omega^4 - 4 \Omega^2 + 4 -\vartheta^2 - \cos^2\left( \frac{q_x}{2}\right) = 0,
    \label{eq:Ch-arg}
\end{equation}~with its roots being:
\begin{equation}
    \Omega = \sqrt{2 \pm \sqrt{\vartheta^2 + \cos^2 \left(\frac{q_x}{2} \right)}}.
    \label{eq:evalues_DL}
\end{equation}
\HA{The dispersion diagram for the diatomic lattice can be depicted by sweeping the non-dimensional wavenumber in the irreducible Brillouin zone, i.e., $q_x \in [0,\pi]$, for both solutions of Equation~(\ref{eq:evalues_DL}), such that the (negative) positive solution corresponds to the (acoustic) optic dispersion branch.}~An alternative method for finding the dispersion relation is to calculate the wavenumber for a given frequency $\Omega$, which is achieved by a simple re-arrangement of the dispersion relation in Equation~(\ref{eq:Ch-arg}):
\begin{equation}
    q_x = \cos^{-1}(\Phi(\Omega)),
    \label{eq:q_driven}
\end{equation}
where
\begin{equation}
    \Phi(\Omega) = 2\left(\Omega^4 - 4\Omega^2 + 4 - \vartheta^2\right)-1.
    \label{eq:Phi_OM}
\end{equation}
Consequently, the resulting wavenumber is generally a complex quantity and its real (imaginary) component represents the frequency (decay rate) of a spatial wave.

Given the discrete nature of the lattice and the presence of elastic foundation, two cutoff frequencies occur and their magnitudes are found by substituting $q_x = 0$ into Equation~(\ref{eq:evalues_DL}):
\begin{equation}
\Omega_{c_\pm} = \sqrt{2\pm\sqrt{1+\vartheta^2}}.
\label{eq:cutoff_freq}
\end{equation}
Additionally, two bandgaps emerge as follows: the \textit{first bandgap} starts \HA{at} $\Omega = 0$ and its upper limit is equal to the lattice's lower cutoff frequency $\Omega_{c_-}$, while the \textit{second bandgap} is sandwiched between the two dispersion branches, in-line with diatomic lattices without elastic foundation, and evaluating Equation~(\ref{eq:evalues_DL}) at $q_x = \pi$ yields its limits:
\begin{equation}
    \Omega_\pm = \sqrt{2\pm|\vartheta|}.
\end{equation}
The maximum attenuation inside both bandgaps can be found from the roots of $\partial_\Omega \Phi~=~0$, i.e., the frequency solutions satisfying the derivative of Equation~(\ref{eq:Phi_OM}) with respect to $\Omega$~\cite{al2017pole}.~The result of this process reveals that the maximum attenuation in the first (second) bandgap occurs at a constant frequency of $\Omega_{\text{max}} =0$ ($\Omega_{\text{max}}=\sqrt{2}$).

\begin{figure*}[]
     \centering
\includegraphics[]{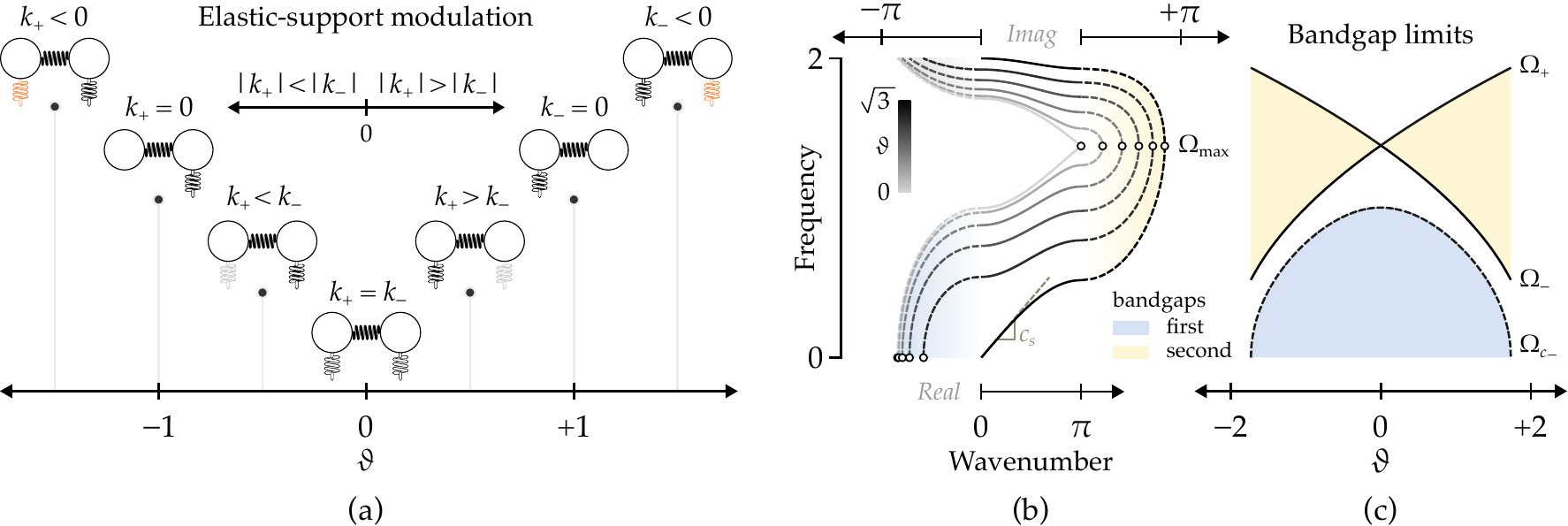}
\caption{(a) Illustration of different combinations of spring constants for the elastic foundation of the diatomic lattice in Figure~\ref{fig:DL_Schematics}\HA{(c)} as the contrast parameter $\vartheta$ changes.~Note that  \HA{the stiffness of one of the springs} becomes negative once $\vartheta$ exceeds a magnitude of unity.~\HA{Given $|\vartheta| > 1$, its sign} indicates which of the springs \HA{$k_+$ or $k_-$} becomes negative. If $\vartheta = 0$, the lattice becomes a simple monatomic lattice with a uniform elastic foundation.~(b) Dispersion relation of \HA{the} elastically-supported diatomic lattice and (c) the corresponding bandgap limits at different values of $\vartheta$.~In subfigure (b), the frequency at the maximum attenuation within bandgaps is indicated by circles within the color-coded first and second bandgap regions.}
\label{fig:Dispersion_Relation_DL}
\end{figure*}

From Equation~(\ref{eq:cutoff_freq}), it is observed that $\Omega_{c_-}$ zeros out if $|\vartheta| = \sqrt{3}$, forcing the first bandgap to close.~Implementing values larger than $|\vartheta| = \sqrt{3}$ returns complex values of frequency $\Omega$, and the system becomes dynamically unstable, explaining the limited range of $\vartheta$ being up to a magnitude of $\sqrt{3}$ (similar to the diatomic lattice discussed by Al~Ba'ba'a~\textit{et~al}~\cite{al2020uncertainty}). This is attributed to the fact that \HA{the stiffness of} one of the elastic supports becomes negative once $\vartheta$'s magnitude becomes larger than unity, which is graphically illustrated in Figure~\ref{fig:Dispersion_Relation_DL}(a).~Interestingly, the diatomic lattice does not immediately lose stability when $\vartheta$ exceeds unity due to the presence of the second positive elastic support, rendering the Hamiltonian $\mathbf{H}$ a positive-definite (or semi-positive-definite) matrix up to $|\vartheta| = \sqrt{3}$. \HAB{It is worthy of note that negative stiffness can be achieved in a variety of ways, including buckled beams~\cite{fulcher2014analytical}, springs hinged with rigid rods~\cite{kapasakalis2021seismic}, or using magnets~\cite{shi2015magnetic,dudek2018negative}.}

To show \HA{the effect of varying $\vartheta$ on the dispersion properties}, Figure~\ref{fig:Dispersion_Relation_DL}(b)~and~(c) show the dispersion relation for different values of $\vartheta$ and the limits for the first and second bandgaps, respectively. As explained earlier, the maximum attenuation in the first and second bandgaps happen at $\Omega_{\text{max}} = 0,\sqrt{2}$, respectively, regardless of $\vartheta$ as inferred from Figure~\ref{fig:Dispersion_Relation_DL}(b). It is also interesting to see that the first bandgap \HA{has a maximum width} at $\vartheta = 0$, at which the second bandgap is closed, while the opposite is true for $|\vartheta| = \sqrt{3}$ (Figure~\ref{fig:Dispersion_Relation_DL}(c)). Notice that, at the limiting \HA{case} of $|\vartheta| = \sqrt{3}$, the acoustic branch of the dispersion relation has a non-zero initial slope (i.e., sonic speed \HA{$c_s$}), which can be shown to have a non-dimensional value of $c_s = 1/4$, following the development in Ref.~\cite{AlBabaa2019DispersionCrystals}.~Otherwise, and for any $|\vartheta| \neq \sqrt{3}$, the initial slope of the acoustic branch is equal to zero (Figure~\ref{fig:Dispersion_Relation_DL}(b)).

\subsection{Finite lattice dynamics}
\HA{
\subsubsection{Equations of motion}}
Studying a finite number of unit cells ($n$) for a diatomic lattice is key to the understanding of natural frequencies within bandgaps, arising from truncation effects \cite{al2017pole,AlBabaa2019DispersionCrystals}. Upon assembling the mass ($\mathbf{M}$) and stiffness ($\mathbf{K}$) matrices for a finite diatomic lattice, the equations of motion can be expressed in the form:
\begin{equation}
    \mathbf{M} \ddot{\mathbf{x}} + \mathbf{K} {\mathbf{x}} = \mathbf{f},
    \label{eq:EOM_DL}
\end{equation}
where $\mathbf{x}$ and $\mathbf{f}$ are displacement and forcing vectors, respectively. \HA{In the absence of external forces, the forcing vector $\mathbf{f}$ becomes a zero vector, pertaining to a free-vibration problem.}~The mass and stiffness matrices are $N \times N$ square matrices that can be of even ($N = 2n$) or odd ($N = 2n+1$) matrix dimension. A diatomic lattice with an even (odd) $N$ has an integer (a non-integer) number of unit cells. Having a lattice with a non-integer number of \HA{unit} cells implies that the last unit cell is incomplete and one of its masses is absent. As such, the lattice at both ends \HA{is} supported by either $k_-$ or $k_+$, making the finite lattice symmetric about its center \cite{al2017pole}. 

\begin{figure*}[h]
     \centering
\includegraphics[]{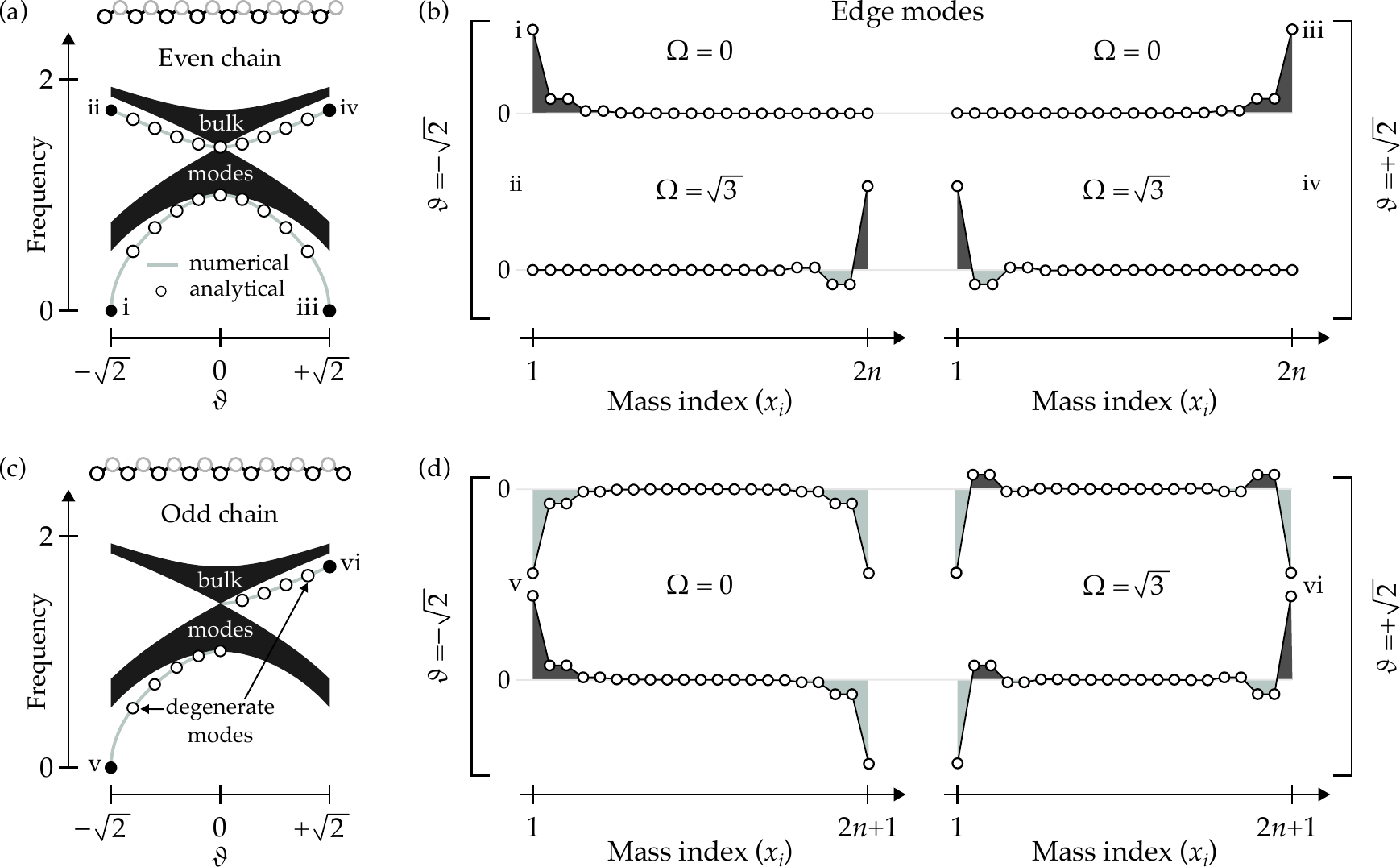}
\caption{Eigenfrequency spectra of an (a) even and (c) odd diatomic chains as a function of the stiffness contrast $\vartheta$. Outside bulk modes and within bandgap regions, there exist edge modes that are computed numerically (green lines) and verified analytically (circles). Modes shapes for edge modes of (b) even and (d) odd chains at $\vartheta = \pm\sqrt{2}$ are also depicted.~The mode shapes are numbered using the numeral i--vi, and their corresponding frequency and $\vartheta$ can be traced to the spectra plots in subfigures (a) and (c).}
     \label{fig:ModeShapes_DL}
\end{figure*}

For notation simplicity, the degrees of freedom for the finite diatomic lattice are labeled as $x_i$, where $i= 1,2, \dots, N$, resulting in the displacement vector:
\begin{equation}
    \mathbf{x} =
    \begin{Bmatrix}
    x_1 & x_2 & \dots & x_N
\end{Bmatrix}^{\text{T}}.
\end{equation} Knowing that odd-numbered (even-numbered) degrees of freedom correspond to masses connected to $k_+$ ($k_-$), a diatomic lattice with free-free boundary conditions has the following mass and stiffness matrices, respectively:
\begin{subequations}
\begin{equation}
    \mathbf{M} = m \mathbf{I}_N,
\end{equation}
\begin{equation}
    \mathbf{K} = 2k \left[ \mathbf{\Psi} + \boldsymbol{\vartheta} \right],
\end{equation}
\label{eq:mat_def_M_K}
\end{subequations}
where $\mathbf{I}_{[ \cdot ]}$ is an identity matrix and its size is indicated by the subscript, and
\begin{subequations}
   \begin{align}
    \underset{N\times N}{\mathbf{\Psi}} & =
    \begin{bmatrix}
    2 + \eta & -\frac{1}{2} & 0 & \cdots & 0 \\
    -\frac{1}{2} & 2 & \ddots & \ddots & \vdots \\
    0 & \ddots & \ddots &  \ddots & 0 \\
    \vdots & \ddots & \ddots &  2 & -\frac{1}{2} \\
    0 & \cdots & 0 &  -\frac{1}{2} & 2+\xi
    \end{bmatrix}, \\
    \underset{N\times N}{\boldsymbol{\vartheta}} & = \mathbf{diag} \begin{bmatrix}
        +\vartheta & -\vartheta & +\vartheta & \cdots & (-1)^{N-1} \vartheta
    \end{bmatrix}.
    \label{eq:psi_th}
\end{align} 
\label{eq:mat_def_DL}
\end{subequations}
The variable $\eta$ ($\xi$) denotes the amount of perturbation in the first (last) element of matrix $\mathbf{\Psi}$ from the rest of \HA{its} diagonal elements, which can be shown to be equal to the off-diagonal elements $a = -1/2$ in the case of free-free boundary conditions.
\HA{\subsubsection{Natural frequencies: Bulk and edge modes}}
Assuming a harmonic solution and applying \HA{a normalization procedure parallel to that} in Sec.~\ref{sec:DL_UC}, the equation~of~motion~(\ref{eq:EOM_DL}) for a free-vibration problem (in conjunction with Eqs.~\ref{eq:mat_def_M_K}~and~\ref{eq:mat_def_DL}) can be written as:
\begin{equation}
\underbrace{
   [\mathbf{\Psi} + \boldsymbol{\vartheta} -\Omega^2 \mathbf{I}_N ]}_{\mathbf{D}(\Omega)} \mathbf{x} = \mathbf{0}.
   \label{eq:dyn_mat_DL}
\end{equation}
As can be inferred from \HA{Equations}~(\ref{eq:mat_def_DL}-\ref{eq:dyn_mat_DL}), the dynamic stiffness matrix $\mathbf{D}(\Omega)$ is a tridiagonal 2-Toeplitz matrix, with alternating diagonal and constant off-diagonal entries. Given the free-free boundary condition as mentioned earlier, the first and last diagonal elements are different from the rest, resulting in a \textit{perturbed} 2-Toeplitz matrix. For \HA{this} type of matrices, an analytical expression for the characteristic equation exists \cite{da2007characteristic}, which has been recently applied to different classes of phononic crystals \cite{al2017pole,al2020uncertainty,bastawrous2022closed}. For a diatomic lattice with an even $N = 2n$, and defining:
\begin{subequations}
\begin{equation}
    d_\pm = 2\pm \vartheta - \Omega^2,
    \label{eq:d_pm_parameter}
\end{equation}
\begin{equation}
    \lambda(\Omega) = d_+ d_-, 
\end{equation}
\text{and},
\begin{equation}
    P_n(\lambda) = a^{2n} U_n \left(\frac{\lambda}{2 a^2} -1 \right),
    \label{eq:quad_tran}
\end{equation}
\end{subequations}
where $U_n(\cdot)$ is the Chebyshev polynomial of the second kind, the characteristic equation for $\mathbf{D}(\Omega)$ is given by \cite{da2007characteristic}:
\begin{equation}
P_n(\lambda) + (\eta d_- + \xi d_+ + \eta \xi  + a^2) P_{n-1}(\lambda) + \eta \xi a^2 P_{n-2}(\lambda) = 0.
    \label{eq:ch_eq_1}
\end{equation}
Interestingly, the dispersion relation in Equation~(\ref{eq:Ch-arg}) can be obtained by equating the argument of the Chebyshev polynomial in \HA{Equation~}(\ref{eq:quad_tran}) to $\cos(q)$, which mandates that
\begin{equation}
U_n\left(\cos (q)\right) = \frac{\sin \left((n+1) q \right)}{\sin (q)}.
\label{eq:Ch-arg-1}
\end{equation} The next goal is to simplify \HA{Equation~}(\ref{eq:ch_eq_1}) and obtain solutions of the non-dimensional wavenumber $q$; following which, the corresponding natural frequencies of the system are found from \HA{Equation~}(\ref{eq:evalues_DL}), i.e., the dispersion relation \cite{al2017pole}.
Making use of \HA{Equations}~(\ref{eq:Ch-arg}), (\ref{eq:quad_tran}),~and~(\ref{eq:Ch-arg-1}) and substituting $\xi = \eta = a$, the characteristic equation~(\ref{eq:ch_eq_1}) reduces to:
\begin{equation}
\frac{\sin(nq)}{\sin(q)} \left( \Omega^4-3\Omega^2+2-\vartheta^2\right) = 0.
    \label{eq:eigenvalues_theta}
\end{equation}
Setting the first factor of \HA{Equation~}(\ref{eq:eigenvalues_theta}) to zero results in the following wavenumber solutions:
\begin{equation}
    q_\ell = \frac{\ell \pi}{n}, \ \ \ell = 1,2, \dots , n-1.
    \label{eq:q_L}
\end{equation}
Subsequently, the natural frequencies can be analytically calculated by substituting the discrete values of the wavenumber in \HA{Equation~}(\ref{eq:q_L}) into the dispersion relation in \HA{Equation~}(\ref{eq:evalues_DL}) and solving for $\Omega$ (one solution per dispersion branch). The resulting solutions satisfy the dispersion relation, thus necessarily lay in the pass-band regions (\HA{referred to here as \textit{bulk modes}}).~Two remaining natural frequencies, which can potentially be within a bandgap, are \HA{computed} by setting the second factor of \HA{Equation~}(\ref{eq:eigenvalues_theta}) to zero, which results in: 
\begin{align}
    \Omega & = \sqrt{\frac{3}{2}\pm \sqrt{\frac{1}{4}+\vartheta^2}}.
    \label{eq:OM_n+1}
\end{align}
The frequency solutions in \HA{Equation~}(\ref{eq:OM_n+1}) correspond to the natural frequencies of a single (unconstrained) free-free unit cell, which immediately follows from setting $n=1$ in \HA{Equation~}(\ref{eq:eigenvalues_theta}). \HA{Further inspection of Equation~(\ref{eq:OM_n+1})} reveals that $\vartheta = 0$ results in $\Omega = 1$ and $\Omega = \sqrt{2}$, which are solutions \HA{that satisfy} the dispersion relation in (\ref{eq:evalues_DL}) at $q_x = 0$ and $q_x = \pi$, respectively.~For $\vartheta \neq 0$, on the other hand, the smaller (larger) solution of Equation~(\ref{eq:OM_n+1}) lives inside the first (second) bandgap, as seen in Figure~\ref{fig:ModeShapes_DL}(a). The corresponding mode shapes for the natural frequencies in \HA{Equation~}(\ref{eq:OM_n+1}), when inside a bandgap, have a localized large amplitude at an \textit{edge} that attenuates in the direction of the lattice's bulk, as seen in Figure~\ref{fig:ModeShapes_DL}(b), thus often named \textit{edge modes} \cite{chen2018topological} (another name for these modes is truncation poles/resonances \cite{al2017pole,AlBabaa2019DispersionCrystals,al2023theory}, as they arise from truncation effects).~The edge mode within the first bandgap, which is also the fundamental mode, zeroes out (i.e., $\Omega = 0$) when $\vartheta=\pm\sqrt{2}$ and coincides with the maximum attenuation of the first bandgap (Figures~\ref{fig:Dispersion_Relation_DL}(b) and~\ref{fig:ModeShapes_DL}(a)).~While unconstrained vibrational systems exhibit zero-frequency modes that correspond to \textit{rigid body modes} (i.e., non-deformable modes), the zero-frequency mode observed here contradicts the non-deformable nature of typical rigid-body modes and, intriguingly, satisfies the characteristics of a regular edge mode as seen in Figure~\ref{fig:ModeShapes_DL}(b).~Emergence of such a zero-frequency edge mode is attributed to the presence of wave attenuation at $\Omega = 0$ as evident from the non-zero imaginary component of the wavenumber (See Figure~\ref{fig:Dispersion_Relation_DL}(b)).~Finally, the second edge mode at $|\vartheta| = \sqrt{2}$ appears \HA{in the second bandgap} at $\Omega = \sqrt{3}$, as inferred from \HA{Equation~}(\ref{eq:OM_n+1}), which also follows the typical characteristics of an edge mode as seen in Figure~\ref{fig:ModeShapes_DL}(b).~Note that $|\vartheta|>\sqrt{2}$ renders the finite diatomic lattice dynamically unstable as it results in a complex frequency solution (See Equation~(\ref{eq:OM_n+1})).

Next, a diatomic lattice with an odd number of masses ($N = 2n+1$) is studied. In such a case, the characteristic equation can be similarly found as follows \cite{al2017pole,da2007characteristic}:
\begin{equation}
    \left(1+\vartheta - \Omega^2 \right) \frac{\sin((n+1)q)}{\sin(q)} + \left(1-\vartheta - \Omega^2 \right) \frac{\sin(nq)}{\sin(q)} = 0.
    \label{eq:D_oddN}
\end{equation}
\HA{
Given that an analytical expression of $q$ from \HA{Equation~}(\ref{eq:D_oddN}) cannot be generally found for $\vartheta \neq 0$, the dispersion relation in \HA{Equation~}(\ref{eq:Ch-arg}) is used in conjunction with \HA{Equation~}(\ref{eq:D_oddN}) to numerically compute solutions of the non-linear system of equations to obtain roots in the form of $(q,\Omega)$ pairs.}  However, an analytical expression for the edge modes, assuming a sufficiently large lattice, can be derived \cite{al2023designing}. For an edge mode, the wavenumber is a complex quantity, i.e., $q = q_{\text{R}} \pm \mathbf{i} q_{\text{I}}$, and the imaginary component signifies the rate of spatial decay (or growth) in the displacement profile. Within attenuation zones, the value of $q_\text{R}$ can be either zero or $\pi$. Implementing $q_\text{I} \neq 0$, and assuming $n \rightarrow \infty$, \HA{Equation~}(\ref{eq:D_oddN}) boils down to:
\begin{equation}
    \left(1+\vartheta - \Omega^2 \right) \text{e}^{q_\text{I}} \pm \left(1-\vartheta - \Omega^2 \right) = 0.
    \label{eq:n_infty_eq_odd}
\end{equation}
Interestingly, a closed-form solution for $q_\text{I}$ can be derived \HA{from Equations~(\ref{eq:q_driven},\ref{eq:Phi_OM})} and can be shown to be:
\begin{equation}
     q_\text{I} = \Im \left(\cos^{-1} \left(\Phi \right) \right) = - \ln \left(\Phi + \sqrt{\Phi^2 - 1} \right).
\end{equation}
Substituting back into \HA{Equation~}(\ref{eq:n_infty_eq_odd}) and performing a few mathematical manipulations, the resulting equation reads:
\begin{equation}
    \left(\Omega^4-3\Omega^2 + 2 -\vartheta^2 \right)^2 = 0.
    \label{eq:edge_modes_DL_odd}
\end{equation}
\HA{Equation~}(\ref{eq:edge_modes_DL_odd}) proves that repeated roots that match those from the even $N$ case in \HA{Equation~(\ref{eq:eigenvalues_theta}) emerge} when a sufficiently large lattice is considered. If $\vartheta<0$ ($\vartheta>0$), the degenerate solutions only satisfy \HA{Equation~}(\ref{eq:D_oddN}) in the first (second) bandgap, with a mode shape localized at both lattice's edges, as illustrated in Figures~\ref{fig:ModeShapes_DL}(c) and (d), respectively. In the following section, the relationship between \HA{the edge modes in the diatomic lattice} and corner modes in two-dimensional mechanical graphene is established. \\

\section{Mechanical graphene}

\subsection{Wave dispersion analysis}
Consider a unit cell of the mechanical graphene as shown in Figure~\ref{fig:DL_Schematics}\HA{(a,b)}; its equations of motion are given by \cite{AlBabaa2020Elastically-supportedInsulators}:
\begin{subequations}
\begin{equation}
    m \ddot{u}_{i,j}+(3k+k_+){u}_{i,j}-k\left(v_{i,j}+ v_{i-1,j} + v_{i,j-1} \right)=0,
\end{equation}
\begin{equation}
m \ddot{v}_{i,j}+(3k+k_-){v}_{i,j}-k\left(u_{i,j}+ u_{i+1,j} + u_{i,j+1} \right)=0.
\end{equation}
\label{eq:Honeycomb_EOM}
\end{subequations}
Owing to graphene's periodicity, the degrees of freedom can be related via two non-dimensional wavenumbers $q_1$ and $q_2$ as follows: $u_{i\pm1,j} = u_{i,j}\text{e}^{\pm\mathbf{i}q_1}$ and $u_{i,j\pm1} = u_{i,j}\text{e}^{\pm\mathbf{i}q_2}$, which similarly apply to $v_{i,j}$ degrees of freedom. The definition of these two non-dimensional wavenumbers in terms of $q_x$ and $q_y$, i.e., the non-dimensional wavenumbers in the $x$ and $y$ directions, are:
\begin{subequations}
\begin{equation}
    q_1 = q_x,
\end{equation}
\begin{equation}
    q_2 = \frac{1}{2}q_x + \frac{\sqrt{3}}{2}q_y.
\end{equation}
\end{subequations}
An identical parameterization of $k_\pm=2k(1\pm\vartheta)$ as in Sec.~\ref{sec:DL_UC} is adopted, yet with a range of $\vartheta \in [-2,2]$ within which the unit-cell's Hamiltonian \HA{of mechanical graphene} remains positive or semi-positive definite.~Assuming harmonic motion, the equations of motion can be compactly represented as in the eigenvalue problem in \HA{Equation~}(\ref{eq:EVP_DL}), with
\begin{equation}
    \mathbf{H} = 
    \begin{bmatrix}
      \frac{5}{2}+\vartheta  & -\frac{1}{2} \left(1+\text{e}^{-\mathbf{i}q_1} +\text{e}^{-\mathbf{i}q_2} \right) \\
      -\frac{1}{2} \left(1+\text{e}^{\mathbf{i}q_1} +\text{e}^{\mathbf{i}q_2} \right)  & \frac{5}{2}-\vartheta
    \end{bmatrix},
    \label{eq:H_HL}
\end{equation}
and, subsequently, the characteristic equation (i.e., dispersion relation) is derived:
\begin{equation}
    \Omega^4 - 5\Omega^2 + 6 - \vartheta^2 -\cos\left(\frac{q_x}{2}\right) \left[\cos\left(\frac{q_x}{2}\right) + \cos\left(\sqrt{3}\frac{ q_y}{2}\right) \right] = 0.
    \label{eq:disp_HC}
\end{equation}
Hence, the dispersion surfaces of mechanical graphene are obtained from the solutions of the dispersion relation in \HA{Equation~}(\ref{eq:disp_HC}):
\begin{equation}
 \Omega = \sqrt{\frac{5}{2}\pm \sqrt{\frac{1}{4}+\vartheta^2 + \cos\left(\frac{q_x}{2}\right) \left[\cos\left(\frac{q_x}{2}\right) + \cos\left(\sqrt{3}\frac{ q_y}{2}\right) \right]}}.
 \label{eq:disp_HC_sol}
\end{equation}
\HA{Analogous} to the diatomic lattice, a first bandgap starting at $\Omega = 0$ opens, owing to the elastic foundation.
In addition, two cutoff frequencies exist, such that the lower cutoff is also the upper limit of the first bandgap. These cutoff frequencies are \HA{the solutions of Equation~}(\ref{eq:disp_HC_sol}) at the $\Gamma$ point (i.e., $q_x = q_y = 0$), which read:
\begin{equation}
    \Omega_{c_\pm} = \sqrt{\frac{5}{2}\pm\sqrt{ \frac{9}{4}+\vartheta^2}}
\end{equation}
Similarly, the limits of the second bandgap can be derived from the dispersion relation in~(\ref{eq:disp_HC_sol}) at a K point (e.g., $q_x = 4\pi/3$ and $q_y = 0$):
\begin{equation}
    \Omega_\pm = \sqrt{\frac{5}{2}\pm |\vartheta|}.
    \label{eq:BGlimits_HC}
\end{equation}

\begin{figure*}[]
     \centering
\includegraphics[]{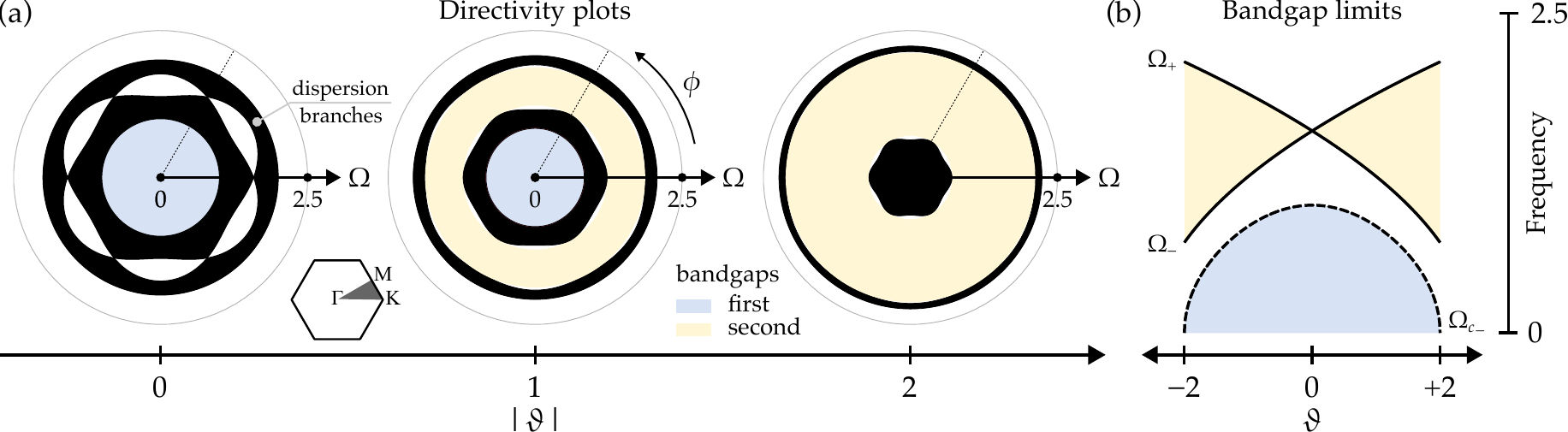}
\caption{(a) Directivity plots for mechanical graphene at different values of the stiffness contrast $\vartheta$, where black areas represent the dispersion branches (i.e., pass band regions), and the first and second bandgaps are color-coded for ease of reference.~(b)~\HA{Evolution} of the limits \HA{of first and second bandgaps} as a function of $\vartheta$.}
     \label{fig:HC_disp}
\end{figure*}

In Figure~\ref{fig:HC_disp}(a), the dispersion relation is depicted for different values of stiffness contrast $\vartheta$ using directivity plots, which is achieved by introducing an angle $\phi = \tan^{-1}(q_y/q_x)$ combining the two non-dimensional wavenumbers $q_x$ and $q_y$. Starting with $\vartheta = 0$, the inversion symmetry of the mechanical graphene remains intact, and the six Dirac points are degenerate with 
no second bandgap opening (i.e., at K and K' points), as observed from Figure~\ref{fig:HC_disp}(a) at a $\phi$ angle of zero and multiples of $60^\text{o}$.~Meanwhile, the first bandgap is open and has its maximum width between the limits $\Omega = 0$ and $\Omega = 1$. At $\vartheta \neq 0$, the inversion symmetry breaks, consequently lifting the degeneracy at the Dirac points and allowing the second bandgap to grow in size with the increase of $\vartheta$ until it peaks at $|\vartheta| = 2$. On the other hand, moving away from $\vartheta = 0$ shrinks the width of the first bandgap until it eventually closes at the limiting case of $|\vartheta| = 2$. This behavior is further verified by the bandgap limits plot as clearly seen in Figure~\ref{fig:HC_disp}(b). 

\subsection{Corner modes}

\begin{figure*}[]
     \centering
\includegraphics[]{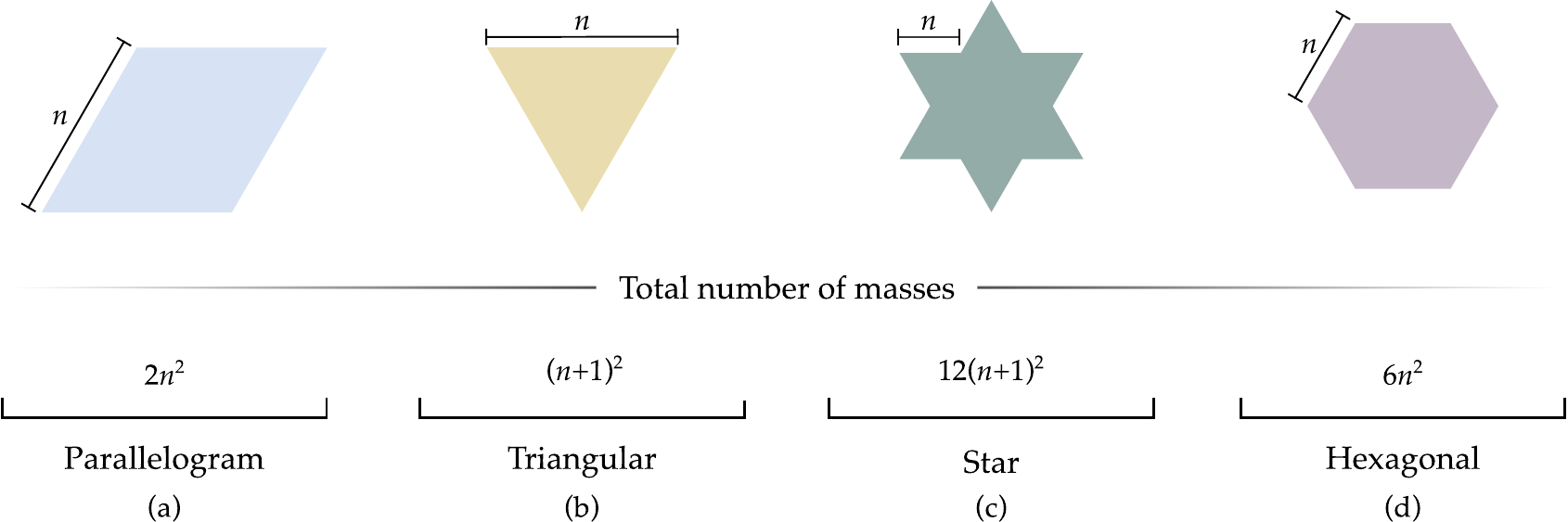}
\caption{Schematics of finite mechanical graphene, showing the different geometries considered, \HA{namely (a) Parallelogram, (b) Triangular, (c) Star, and (d) Hexagonal shapes. The total number of degrees of freedom in each case is indicated and it is a function of the variable $n$, representing} the number of complete unit cells on the equal-sided structures.}
\label{fig:Shapes_HC}
\end{figure*}

\begin{figure}[]
     \centering
\includegraphics[]{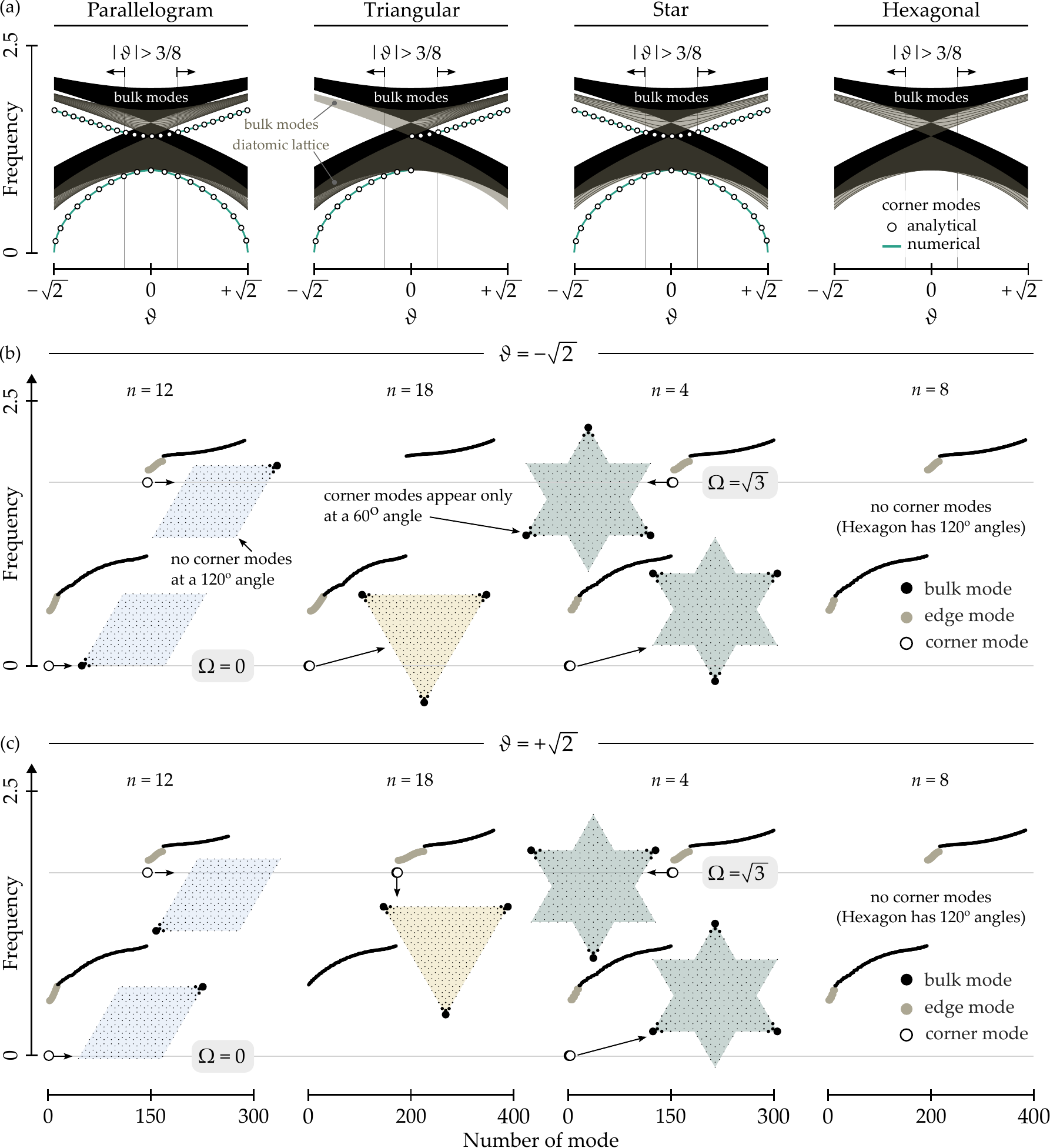}
\caption{\HA{(a)} Eigenfrequency spectra of the different shapes of mechanical graphene at a swept range of $\vartheta$.~Bulk modes are represented by the black regions.~The black lines in the figure are classified as edge modes, depicted within the orange shading signifying the regions of bulk modes of a diatomic lattice realization of the mechanical graphene.~The rest of the lines (depicted in green) represent corner modes, which are verified via the analytical model (circles) corresponding to the diatomic-lattice's edge modes in Equation~(\ref{eq:OM_n+1}).~Within the second bandgap, corner modes only appear when $|\vartheta|>3/8$.~(b,c) Eigenfrequency spectra of the different shapes of mechanical graphene at $\vartheta = \pm\sqrt{2}$, and the mode shapes of corner modes (if any) at $\Omega = 0,\sqrt{3}$ are shown.~Note that no corner modes emerge in the case of the hexagonal shape due to the lack of $60^\text{o}$ angle corners.~\HA{Simulations are conducted for the different shapes of mechanical graphene according to the indicated} number of complete unit cells $n$ on \HA{their structural} edges (\HA{See Figure~\ref{fig:Shapes_HC} for details on the definition of $n$ for the different cases)}.}
\label{fig:Modes_HC2}
\end{figure}

A finite mechanical graphene can be constructed in a variety of ways, having a parallelogram, triangular, star, or hexagonal shape with equal sides as illustrated in Figure~\ref{fig:DL_Schematics}\HA{(c)}.~The total number of degrees of freedom is characterized via the number of complete unit cells $n$ on the diatomic lattices comprising the \HA{structural} edges, which are graphically summarized in Figure~\ref{fig:Shapes_HC}.~All cases have free boundary conditions at the structure's equal sides, and the pattern of the elastic supports is placed such that the first mass in the bottom row is connected to $k_+$, starting from the left. 

\HAB{Using an in-house code developed via MATLAB,} eigenfrequency spectra are calculated for the parallelogram, triangular, star, and hexagonal shapes of mechanical graphene with $n= 12$, $n = 18$, $n = 4$ and $n = 8$ for $\vartheta \in [-\sqrt{2},+\sqrt{2}]$, as seen in Figure~\ref{fig:Modes_HC2}(a). Corner modes only appear for the three first cases, and are discerned as isolated lines within a bandgap and further verified from their mode shapes. Interestingly, the frequency at which corner modes materialize, shown as green lines, is exactly \HA{equal} to the edge states of a single strip of mechanical graphene (i.e., the diatomic lattice) in Equation(\ref{eq:OM_n+1}), depicted as circles in the figure. For the first bandgap, corner mode(s) appears at any non-zero value of $\vartheta$, which, interestingly, is (are) the fundamental mode(s) of the entire system.~Only at a value of $|\vartheta| = \sqrt{2}$ such corner modes assume a magnitude of zero, in agreement with the edge modes predicted in the diatomic lattice using \HA{Equation~}(\ref{eq:OM_n+1}). The corner modes in the second bandgap, on the other hand, only appear if the modulation is sufficiently large and $|\vartheta|>3/8$. \HA{Such a range of contrast parameter $\vartheta$ is deduced} by examining at what condition the edge states in Equation~(\ref{eq:OM_n+1}) for a diatomic lattice fall within the bandgap limits of mechanical graphene in Equation~(\ref{eq:BGlimits_HC}). A final note on the frequency spectra in Figure~\ref{fig:Modes_HC2}\HA{(a)} is that edge modes in mechanical graphene (\HA{i.e., modes with localization at the structural boundaries, while the bulk being nearly still}) coincide with the range of bulk modes of a diatomic lattice \HA{constituted from a single strip of mechanical graphene}, as highlighted in light orange in the figure.~This is of interest, as it implies that all edge modes in two-dimensional mechanical graphene \HA{are} also potential natural frequencies within pass bands of \HA{its one-dimensional realization, the diatomic lattice}. 

To further understand corner-mode emergence \HA{in the different shapes of mechanical graphene}, the natural-frequency spectra at $\vartheta = \pm \sqrt{2}$ and corresponding mode shapes for corner modes (if any) are depicted in Figure~\ref{fig:Modes_HC2}\HA{(b,c)} for all structural shapes. As seen from the mode shapes and natural frequency distribution, several observations are made:
\begin{enumerate}
    \item For $\vartheta = \pm \sqrt{2}$, corner modes may appear at the frequencies $\Omega = 0$ and $\Omega = \sqrt{3}$, located within the first and second bandgap, respectively, depending on the overall shape of mechanical graphene. 
    \item Regardless of the frequency at which a corner mode originates, its mode shape always has a localization at a $60^\text{o}$-shaped corner, explaining the lack of corner modes in the case of \HA{hexagonal-shaped mechanical graphene}. The location of corner-mode localization depends on the stiffness of the spring supporting the corner, such that a corner mode within the first (second) bandgap localizes vibrational energy at a corner supported by a smaller (larger) stiffness. Note that at $\vartheta = \pm \sqrt{2}$, the smaller elastic support also has a negative stiffness constant.
    \item The total number of corner modes appearing in both bandgaps is equal to the total number of corners with $60^\text{o}$ shape.~For instance, a star-shaped mechanical graphene has six corners with $60^\text{o}$ angle, \HA{enabling} a total of six corner modes. \HA{These corner modes may appear in a single bandgap or split evenly between the first and second bandgaps, as detailed next.} 
    \item A mechanical graphene with a parallelogram shape exhibits a single corner mode per bandgap and a total of two corner modes, \HA{given} the two $60^\text{o}$ corners.~Analogously, the star-shaped mechanical graphene has three corner modes per bandgap, owing to the six $60^\text{o}$ corners.
    \item For a parallelogram and star shapes, flipping the sign of $\vartheta$ results in changing the location of corner modes'~localization at a given frequency.~This behavior is attributed to the fact that the elastic supports of smaller and larger stiffnesses swap with changing $\vartheta$'s sign.
    \item A triangular-shaped mechanical graphene is a spacial case.~At a given sign of $\vartheta$, the corner modes appear only in one of the bandgaps, as all corners are connected to \HA{matching} elastic supports.~This phenomenon is inline with \HA{the emergence of} edge modes in a diatomic lattice with an odd number of degrees of freedom as seen in Figure~\ref{fig:ModeShapes_DL}(c).~Indeed, triangular-shaped mechanical graphene shall always have an edge mimicking an odd diatomic chain. Therefore, corners with \HA{elastic supports of smaller (larger) stiffness} result in corner modes only within the first (second) bandgap.
\end{enumerate}

\begin{figure*}[]
     \centering
\includegraphics[]{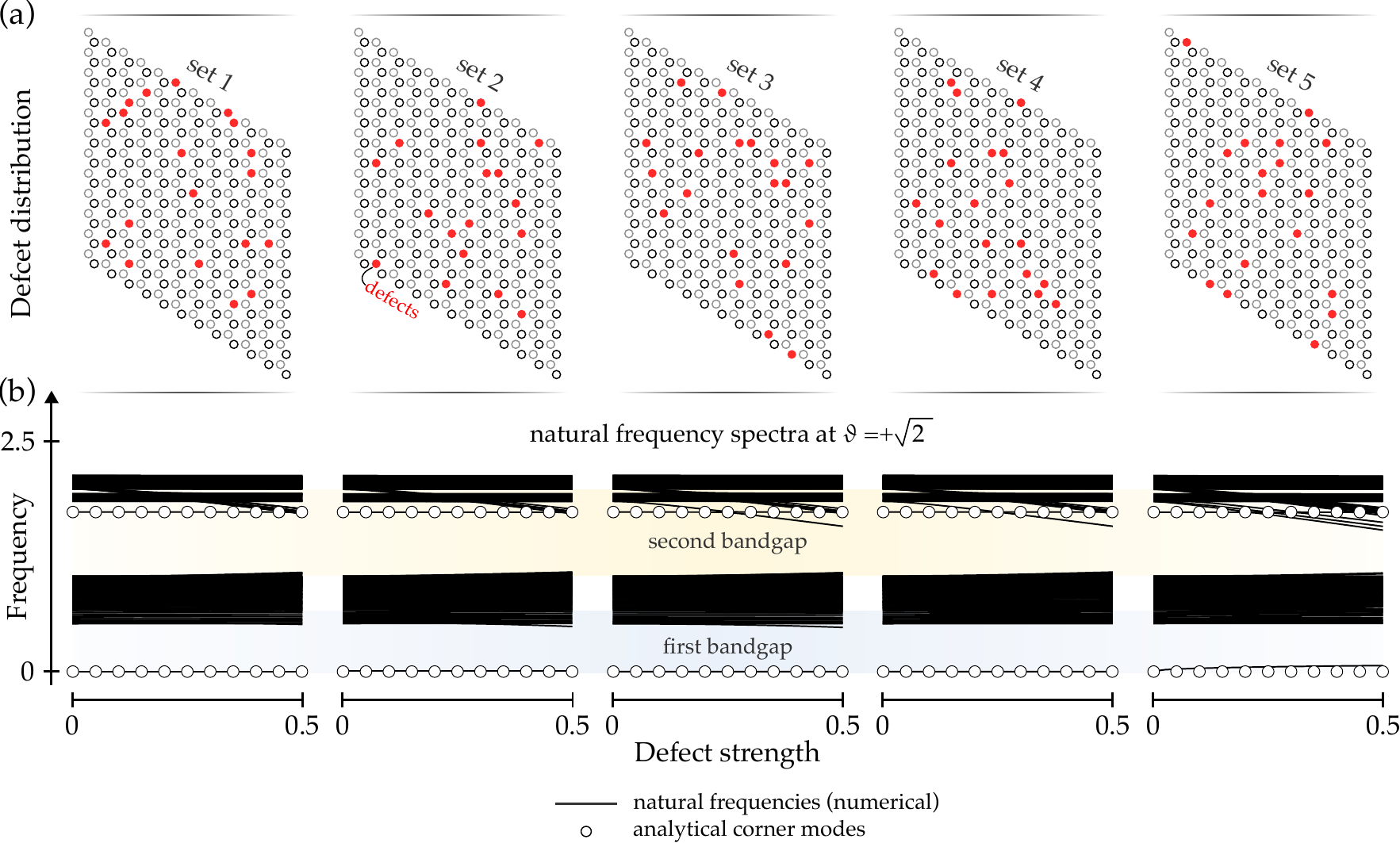}
\caption{\HAB{(a) Schematics of finite parallelogram-shaped mechanical graphene, showing different defect location sets with a total of twenty defects, randomly distributed throughout the structure. Using an identical parameter set for parallelogram-shaped mechanical graphene in Figure~\ref{fig:Modes_HC2}(c), the lower (upper) $60^\text{o}$ corner is supported by a positive (negative) elastic-foundation spring constant.~(b) Natural-frequency spectra of each set is computed to test the effect of defect location and strength, demonstrating that only the last defect set shifts the zero-frequency corner mode.~This is because a defect is near the upper $60^\text{o}$ angle corner (with negative elastic support), at which the zero-frequency corner mode is localized.}}
\label{fig:Defect_HC}
\end{figure*}

\HAB{Finally, the effect of defects on the zero-frequency corner modes (and natural-frequency spectra in general) is briefly discussed. The elastic-foundation springs are assumed to be defected and the defect strength is expressed by the factor $r \in [0,0.5]$, such that the defected sites are of stiffness $\hat{k}_\pm = 2k(1-r)(1\pm \vartheta)$.~Therefore, a zero value of $r$ results in the pristine case (with no defects), and the level of the defect increases with a growing value of $r$.

To perform the test, twenty, randomly placed defects, are imposed on a parallelogram-shaped mechanical graphene with $\vartheta = + \sqrt{2}$ and $n = 12$, identical to the parameters used in Figure~\ref{fig:Modes_HC2}(c). Note that the lower (upper) $60^\text{o}$ corner is supported by a positive (negative) elastic-foundation spring constant based on the chosen parameters. Five different sets of defect locations are generated, which are shown in Figure~\ref{fig:Defect_HC}(a), with their corresponding natural-frequency spectra depicted in Figure~\ref{fig:Defect_HC}(b). It can be seen that both the defect placement and its strength affect the resulting natural frequency spectra, and more modes fall within both bandgaps with increasing defect strength. For the first four cases, the effect on the zero-frequency corner mode seems minimal and they appear fairly robust. However, a defect near a corner has a higher effect on perturbing the frequency at which the corner mode emerges, which clearly evident from the fifth set in the rightmost panel of Figure~\ref{fig:Defect_HC}. As the defect is near the upper corner where the negative stiffness is placed and the zero-frequency corner mode is localized, the first natural frequency (i.e., the zero-frequency corner mode) deviates further from zero as the defect grows stronger, which is not the case for the first four cases with defects well within the bulk of the structure.}

\section{Concluding remarks}
In this paper, a mechanical graphene built from a honeycomb tessellation of elastically-supported masses is proposed with the capability of exhibiting corner modes at zero frequency.~To enable zero-frequency corner modes, alternating positive-negative elastic supports for the two sites of mechanical-graphene's unit cell are imperative, and their stiffnesses are carefully calculated to ensure dynamical stability.~Examining the dynamics of a diatomic chain, made from a single strip of the mechanical graphene \HA{with free boundaries}, has been instrumental in unraveling the origins of the corner modes.~It has been shown that corner modes emerge at an identical frequency to that of edge modes in the diatomic lattice, provided that the structural shape of mechanical graphene has $60^\text{o}$ corners.~Four different mechanical graphenes with parallelogram, triangular, star, and hexagonal shapes have been investigated.~It is proven that the total number of corner modes (appearing in one or both bandgaps) is equal to the number of $60^\text{o}$-shaped corners, explaining the lack of corner modes for the hexagonal-shaped mechanical graphene.~\HAB{Brief analysis on the effect of defect location and strength on zero-frequency corner modes has been also conducted, showing their robustness against defects that are well within the structural bulk.} Although such zero-frequency corner modes are fully explained here through the lens of vibration and analytical characteristic equations, further investigations are necessary to determine whether these modes are topologically protected or not.

\section*{Acknowledgment}
The author expresses his gratitude to Prof. Harold Park from Boston University and Dr. D. Zeb Rocklin from Georgia Institute of Technology for useful discussions.

%\section*{Conflicts of interest}
%The author has no conflicts to disclose.

%\section*{Data Availability Statement}
%The data that support the findings of this study are available from the corresponding author upon reasonable request.

\section*{References}
\begin{multicols}{2}
\footnotesize
\printbibliography[heading=none]
\end{multicols}
\break

\end{document}